\documentclass[fleqn,usenatbib]{mnras}

\usepackage[T1]{fontenc}

\usepackage{graphicx}	
\usepackage{amsmath}	
\usepackage{amssymb}	

\usepackage{newtxtext,newtxmath}

\newcommand\gaia{\textit{Gaia\/}~}
\newcommand{\textcode}[1]{\texttt{#1}}
\DeclareRobustCommand{\VAN}[3]{#2}
\let\VANthebibliography\thebibliography
\def\thebibliography{\DeclareRobustCommand{\VAN}[3]{##3}\VANthebibliography}
\newcommand{\changed}[1]{#1}  

\title[Gaia DR3 and nearby galaxies]
{Gaia DR3 and nearby galaxies: where do foregrounds matter?}

\author[Barmby]{
P. Barmby$^{1}$\thanks{E-mail: pbarmby@uwo.ca}
\\
$^{1}$Department of Physics \& Astronomy and Institute for Earth and Space Exploration,
Western University, London, Canada N6A 3K7
}

\date{Accepted XXX. Received YYY; in original form ZZZ}

\pubyear{2022}

\begin{document}
\label{firstpage}
\pagerange{\pageref{firstpage}--\pageref{lastpage}}
\maketitle

\begin{abstract}
Nearby galaxies provide populations of stellar and non-stellar sources at a common distance and in quantifiable environments. 
All are observed through the Milky Way foreground, with varying degrees of contamination that depend on observed Galactic latitude and the distance and size of the target galaxy.
This work uses \changed{\gaia Data Release 3 (DR3)} to identify foreground sources via astrometric measurements and thus quantify foreground contamination for a large sample of nearby galaxies.
There are approximately half a million \gaia sources in the directions of \changed{1401} galaxies listed in the Local Volume Galaxy catalogue ($D<11$~Mpc), excluding the largest Local Group galaxies.
About two thirds of the \gaia sources have astrometric properties consistent with foreground sources; these sources are brighter, redder, and less centrally-concentrated than non-foreground sources.
Averaged over galaxies, foreground sources make up 50 per cent of \gaia sources at  projected radius  $r_{50}=1.06 a_{26}$, where $a_{26}$ is the angular diameter at the $B=26.5$ isophote.
Foreground sources make up 50 per cent of \gaia sources at apparent magnitude $m_{G, 50}=20.50$.
This limit corresponds to  the tip of the red giant branch  absolute magnitude at $D=450$~kpc, and to the globular cluster luminosity function peak absolute magnitude at 5~Mpc.
\gaia data provide a powerful tool for removing foreground contamination in stellar population studies of nearby galaxies, although
\changed{\gaia foreground removal will be incomplete} beyond distances of 5~Mpc. 

\end{abstract}




\section{Introduction}
\label{sec:intro}

Observations of nearby external galaxies form the basis for understanding of the diversity of present-day galaxy properties and their variation with internal and external environment.
They complement studies of the Milky Way, where detailed studies of individual constituent objects are possible but measurement of distances to these objects and of integrated galaxy properties are difficult.
Unlike the situation for distant galaxies, with nearby galaxies one cannot simply choose to study only sources at high Galactic latitudes and expect to get a complete picture of the population.
Nearby galaxies are viewed through the Milky Way and their study is complicated by extinction from the Galactic interstellar medium and contamination from foreground Galactic stellar populations.

The \gaia mission \citep{gaia_collab2016,gaiadr3} provides a new way to characterise the Galactic foreground stellar populations in front of nearby galaxies, and in some cases the contents of the galaxies themselves.
The power of the \gaia data is found in their uniformity across the sky and well-characterised uncertainties.
All-sky maps of \gaia source density demonstrate that the \gaia data contain many extragalactic sources: see, for example, figure~3 of \citet{boubert2020}, in which the Magellanic Clouds and M31 as well as several Galactic star clusters are clearly visible off the plane of the Milky Way. 
The faintest sources with reliable parallax measurements have distances 
less than a few kiloparsecs, so a parallax measurement for such a source is an unambiguous indication that an object cannot be a member of even the nearest external galaxies.
Reliable measurement of stellar proper motions extends to more distant stars  and can be used to associate a star with either the Milky Way or a nearby galaxy \citep[e.g.][]{mcconnachie2021, battaglia2021}.
Non-detection of parallax or proper motion is a necessary but not sufficient indicator that an object is located beyond the Milky Way. 
For example, in searching for globular cluster candidates around NGC~1399 and NGC~3115, \cite{Buzzo2022} excluded sources whose \gaia proper motions were inconsistent with zero.

As well as using \gaia data to identify foreground contaminants, several recent works have made use of \gaia data to study the stellar contents of nearby galaxies themselves. 
Some examples include \citet{grady2021}, who mapped structures in the Magellanic Clouds using photometric estimates of individual stellar metallicities, \citet{mcconnachie2021} who used \gaia proper motions to investigate the orbits of isolated Local Group dwarf galaxies, \citet{huang2021} who searched for globular clusters associated with Milky Way dwarf satellites and \citet{Qi2022} who studied stellar proper motions in the outskirts of Milky Way dwarf satellites.
\gaia observations have been used to measure average proper motions of Local Group galaxies such as M31 and M33 \citep{van_der_marel2019} and to identify individual supergiant stars in M31 and M33 \citep{massey2021, Maravelias2022}, the LMC  \citep{yang2021a} and NGC~6822 \citep{yang2021b}.
As one of the nearest large galaxies outside the Local Group,  NGC~5128 is an attractive target for \gaia studies.
\citet{voggel2020} and \citet{hughes2021} used \gaia observations to identify new candidate ultra-compact dwarf galaxies (UCDs) and luminous globular clusters in this galaxy's halo. 
\citet{voggel2020} estimated that \gaia studies can identify UCDs out to distances of approximately 25~Mpc.
Several pre-launch predictions on \gaia detection of much more distant, unresolved galaxies \citep{desouza2014,debruijne2015} have not yet been followed-up with \gaia studies.

The purpose of this work is to inventory the contents of \changed{\gaia Data Release 3 (DR3)} in the directions of nearby galaxies.
With a a well-defined method for determining which sources have parallaxes and/or proper motions inconsistent with a target galaxy, the number, spatial distribution, and apparent magnitude distributions can be compared  between foreground and non-foreground sources on a per-galaxy basis.
The apparent magnitude distributions can also be compared to relevant luminosities of sources within the target galaxies.
The overall research question is ``where do foregrounds matter?" as determined by projected galactocentric radius within a galaxy \changed{field}, position within the colour-magnitude diagram, and overall galaxy properties.

\section{Data and Methods}
\label{sec:data}

The characteristics of \gaia DR3 are described in detail by \citet{gaiadr3}.
Briefly, \gaia DR3 contains $1.8\times10^9$ objects detected down to limiting magnitudes of $G\sim 21$, with five- or six-parameter astrometry (position, proper motion, parallax) for approximately two thirds of these.
The \gaia data are still being collected and processed, and as such the precision and accuracy vary across the sky depending on how many times a particular area has been scanned \citep{lindegren2021}.
As a rough indication, typical \gaia uncertainties are 0.5~mas~yr$^{-1}$ in proper motion and 0.5~mas in parallax at $G\approx 20$, with uncertainties roughly a factor of 20 smaller for $G<15$.
Most stars in even the nearest Milky Way satellites are too faint for accurate \gaia parallax measurements, but individual proper motions are measurable for the brightest stars in galaxies throughout the Local Group \citep[e.g.][]{mcv2020,battaglia2021}.

As a sample of nearby galaxies, I chose the Local Volume Galaxy sample of \citet[][hereafter LVG]{karachentsev2013}\footnote{%
This catalogue has been updated since the above publication; the edition dated \changed{2022 June} was used, downloaded from \url{http://www.sao.ru/lv/lvgdb/}.}.
The catalogue contains \changed{1421} objects that are located within 11~Mpc of the Milky Way or that have measured radial velocities $<600$ km~s$^{-1}$ with respect to the Local Group centroid.
It does not include the Virgo cluster but does include over a dozen groups similar in size and population to the Local Group \citep{karachentsev2013}.
The majority of galaxies within the volume are classified as spheroidal dwarfs, with $L/L_{\rm MW}<10^{-4}$, typical radius $\sim 2$~kpc, and typical mass within the Holmberg radius $\sim 10^{8}$~M$_{\odot}$.
Because the nearest large galaxies have already been the subject of individual investigations with {\em Gaia}, I removed M31, M33, Sagittarius dSph, the Large and Small Magellanic Clouds, and the Milky Way itself, from the present study. 
Also removed were M32, because it is seen in projection against the M31 disk and will thus be heavily contaminated, and the large-angular-diameter dwarfs Antlia~2, Crater~2, and B\"ootes~III \citep{torrealba2019,torrealba2016,grillmair2009}, which are expected to be completely dominated by foreground sources. 
\changed{Ten} galaxies listed in the LVG without visible-light size measurements were also removed; these are poorly-studied objects detected only by their 21~cm  atomic hydrogen emission.
The final source list contains \changed{1401} galaxies. 

There is no one way to define the size of a galaxy; for the purposes of this study  a consistent measure is needed.
As a size indicator to define a \gaia search radius, I used
the LVG-reported values of $a_{26}$ (the angular diameter at the $B=26.5$ isophote) as the {\em radius} of a search cone.
Extending the search cone to double the isophotal radius was done with the intention of reaching foreground-dominated galactocentric distances.

\changed{
Foreground sources are identified based on their DR3 parallaxes ($\varpi$) and proper motions $\mu_{\alpha^*, \delta}$ and the corresponding uncertainties in those quantities and the covariances between them ($\sigma_{\varpi}, \sigma_{\mu, \alpha^*}, \sigma_{\mu, \delta}, {\rm Cov}(\mu)$).%
Sources are identified as foreground at a level of significance $n\sigma$ if they satisfy either of two criteria similar to those used by \citet{klioner2022}, either significant non-zero parallax:
\begin{equation}
\left| \frac{\varpi + 0.017}{\sigma_{\varpi}}\right| >n
\label{eq:parallax_crit}
\end{equation}
where 0.017~mas is the median parallax zeropoint determined by \citet{lindegren2021}, or significant non-zero proper motion:
\begin{equation}\label{eqn:pm} 
\left[\begin{array}{ll} \mu_{\alpha^*} & \mu_{\delta} \end{array} \right] {\rm Cov}(\mu)^{-1} 
\left[ \begin{array}{l} \mu_{\alpha^*} \\ \mu_{\delta} \end{array} \right] > n^2.
\end{equation}
\autoref{eqn:pm} is implemented following \citet{klioner2022} as 
}
\begin{verbatim}(power(pmra/pmra_error,2) 
+power(pmdec/pmdec_error,2) 
-2*pmra_pmdec_corr 
  *pmra/pmra_error 
  *pmdec/pmdec_error)
/(1-power(pmra_pmdec_corr,2))> power(n,2)
\end{verbatim}

\changed{
The analysis here differs from that of \citet{klioner2022}, who were concerned with identifying distant sources to define the celestial reference frame.  As discussed in \autoref{sec:intro}, some galaxies with the Local Group are near enough for their stars to have proper motions measurable with \textit{Gaia}. For these galaxies I add an additional classification criterion: for a source to be classified as foreground, it must have proper motions incompatible with the parent galaxy's systemic proper motion as listed by \citet{battaglia2021}:
\begin{equation}
(\mu_{\alpha^*}\pm n\sigma_{\mu {\alpha^*}}) \notin (\mu_{\alpha^*, {\rm glx}}\pm n\sigma_{\mu_{\alpha^*, {\rm glx}}})
\label{eq:pmra}
\end{equation}
and/or 
\begin{equation}
(\mu_{ \delta}\pm n\sigma_{\mu { \delta}}) \notin (\mu_{ \delta, {\rm glx}}\pm n\sigma_{\mu_{ \delta, {\rm glx}}})
\label{eq:pmdec}
\end{equation}
Sources without astrometric measurements are classified as non-foreground. In the analysis that follows,  $n=2,3,5,10$  are  considered.
}

Unlike other \gaia analyses where the goal is to select a high-quality sample with a high membership probability for a specific galaxy, here the goal is to achieve a broad characterization of the DR3 contents around nearby galaxies.
As such, I did not apply quality flags when selecting \gaia sources, or attempt to use them to separate background galaxies from members of the target nearby galaxies.
Previous studies of nearby galaxies provide additional information that could be used to determine whether a source is in the foreground, in the target galaxy, or in the background: for example, radial velocity measurements, high-resolution imaging, and multi-wavelength detection. 
Such information is valuable but also wildly heterogeneous across galaxies; in the interests of consistency, it was not used here. 
As described by \citet{brown2021}, \gaia measurements in regions of high projected stellar densities are potentially affected by crowding. 
This effect is a potential shortcoming of the present analysis; correcting for it is beyond the scope of this work.
Some of the very brightest Galactic stars are not (yet) included in the \gaia processing or DR3 database \citep{brown2021}, but as these are rare their omission should not significantly affect the results.

After identifying foreground and non-foreground \gaia sources in the direction of each galaxy \changed{field}, distributions of foreground and total source densities in both  projected galactocentric distance and magnitude are compared.
To ensure meaningful measurements that are not dominated by small number statistics, these distributions are computed only for galaxies with 50 or more \gaia sources.
Kernel density estimates of the two distributions are performed separately, then used to compute the ratio of foreground to total sources which is then fit to a polynomial that is used to compute the radius or apparent magnitude at which 50  per cent of the sources are foreground, denoted as $r_{50}$ and $m_{50}$, respectively.

\subsection{Case study: NGC 2403}

\begin{figure}
\includegraphics[width=\columnwidth]{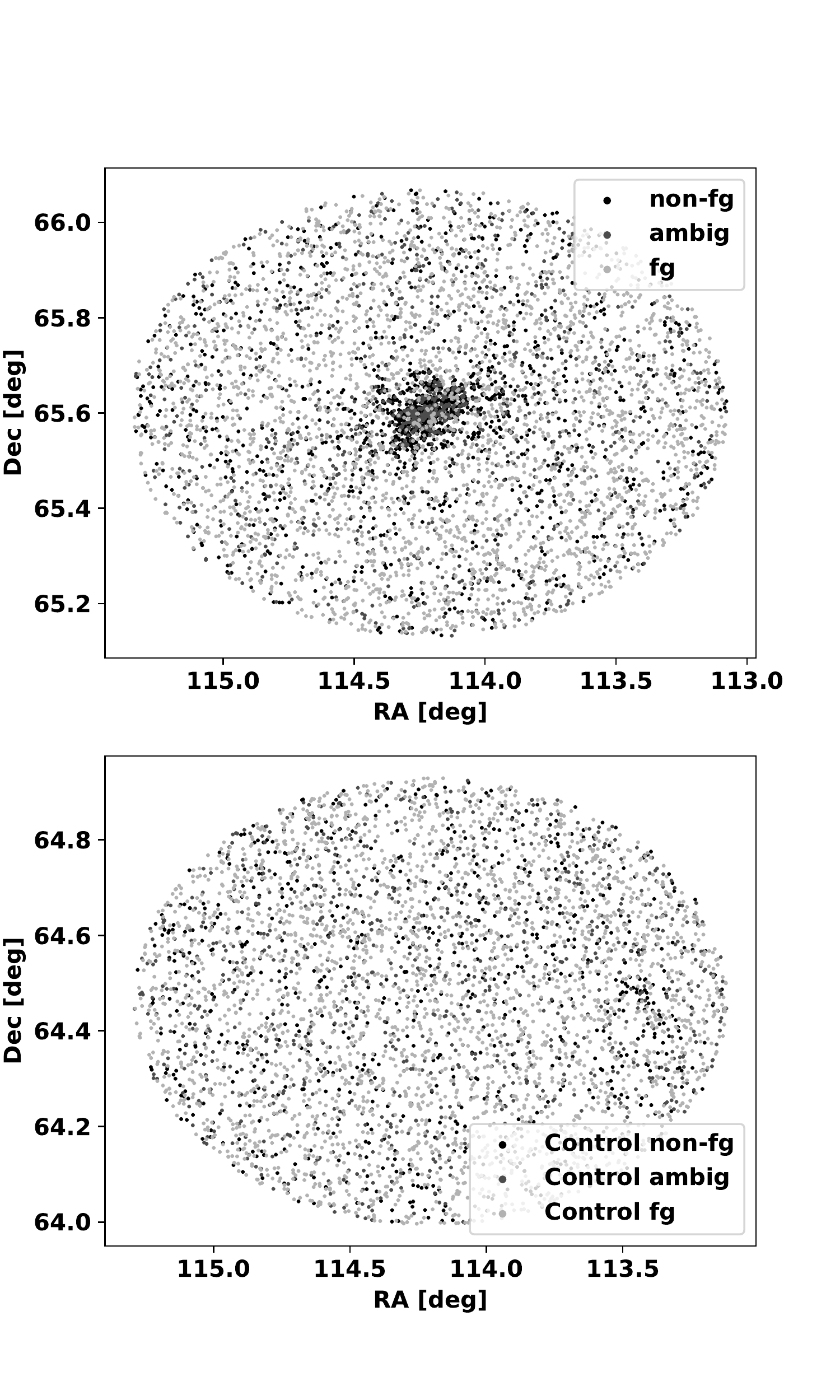}
\caption{
Spatial location of foreground, ambiguous, and non-foreground  \gaia sources within a 28.1~arcmin radius of NGC~2403 (upper panel) and a \changed{nearby} control field (lower panel).
Non-foreground sources are more concentrated toward the centre of the galaxy field but not the control field.}
\label{fig:example_spatial}
\end{figure}

\begin{figure}
\includegraphics[width=\columnwidth]{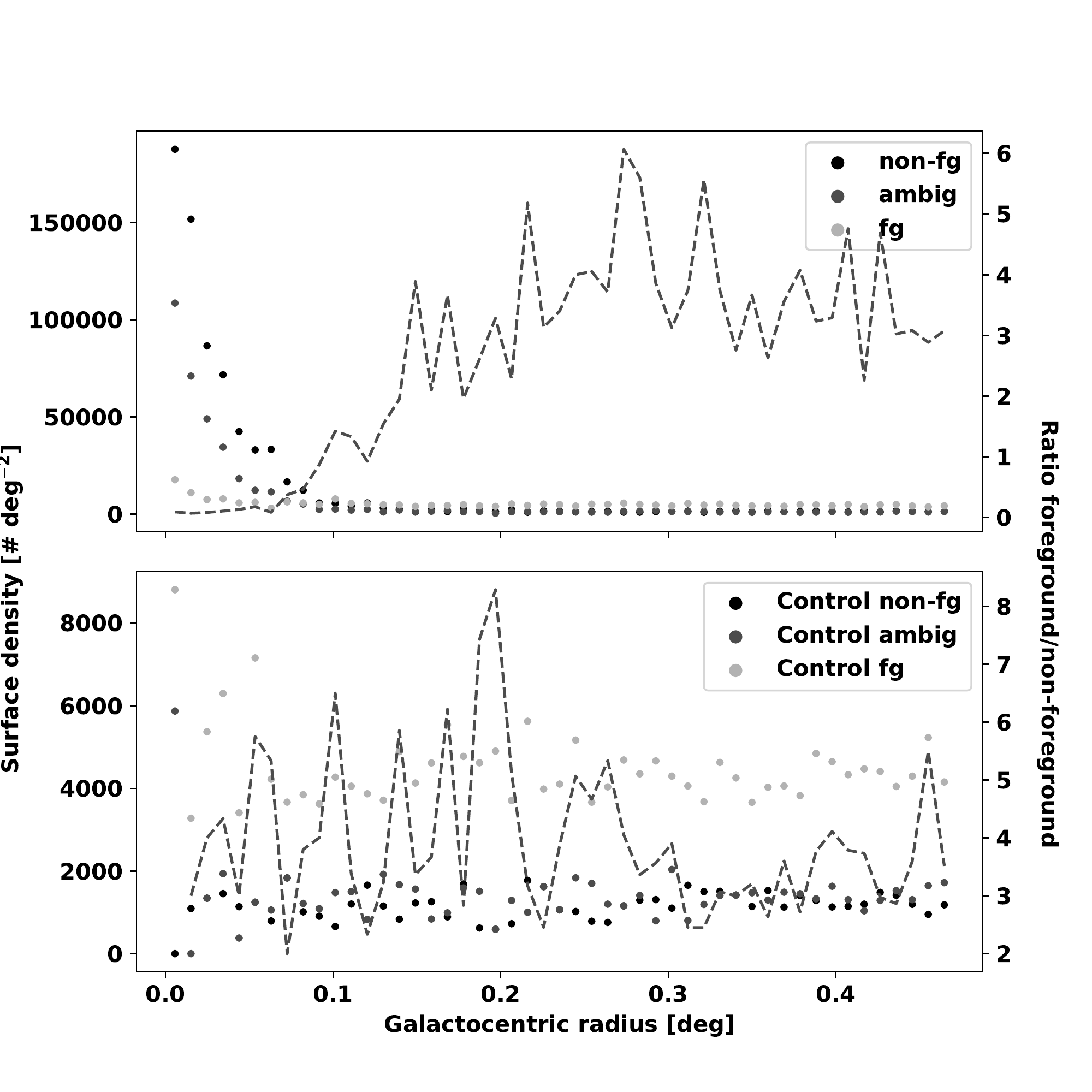}
\caption{
Left axis: projected number densities (left axis) versus projected galactocentric radius for \gaia sources within a 28.1~arcmin radius of NGC~2403 and a control field.
Black points: non-foreground sources, \changed{dark grey points: ambiguous sources, light grey points: foreground sources.}
Right axis and dashed grey line: ratio of foreground/non-foreground source densities. 
}
\label{fig:example_radial}
\end{figure}

\begin{figure}
\includegraphics[width=\columnwidth]{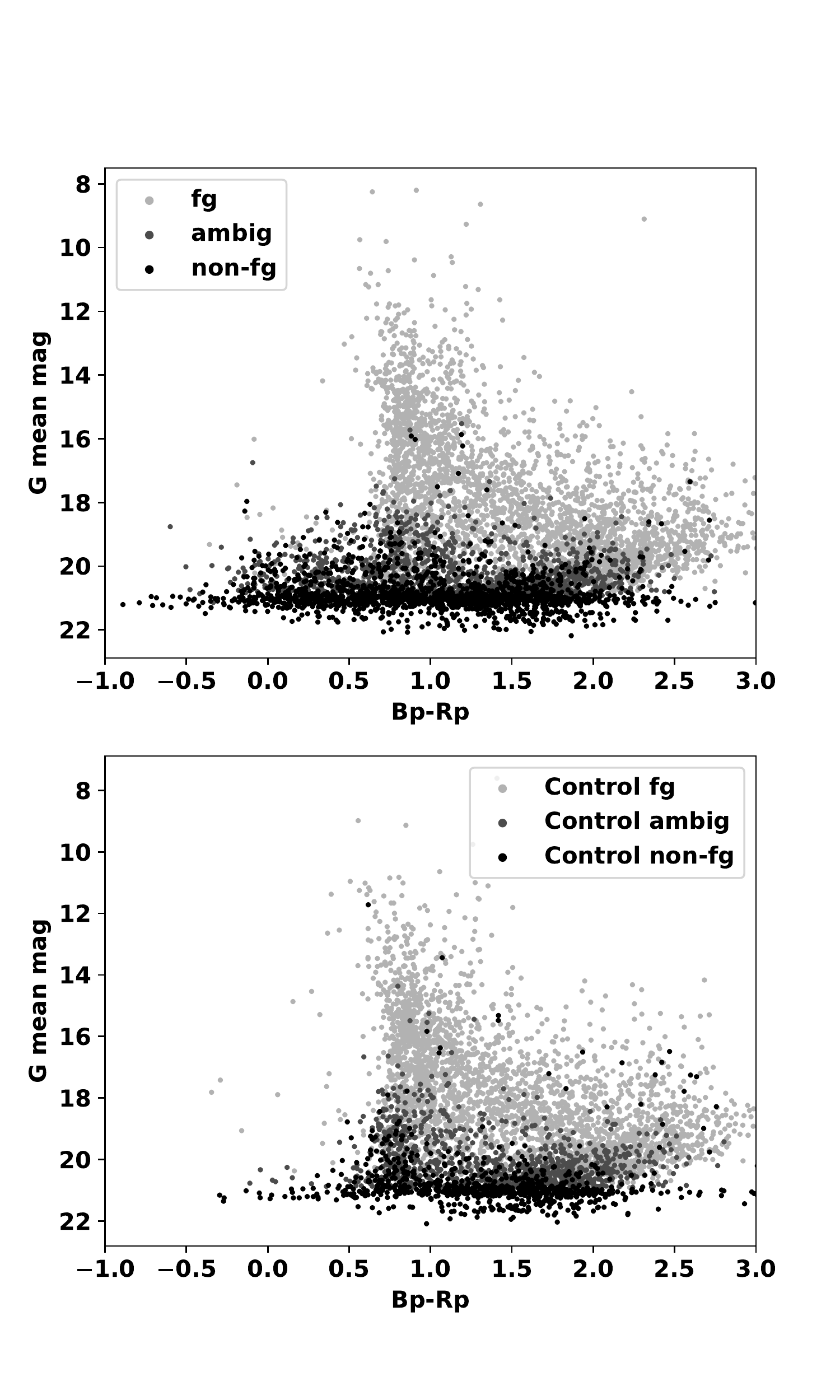}
\caption{
Colour-magnitude diagram of \changed{foreground (light grey), ambiguous (dark grey)} and non-foreground (black) \gaia sources within a 28.1~arcmin radius of NGC~2403 (upper panel) and a control field (lower panel).
Measurements are not corrected for extinction.
}
\label{fig:example_cmd}
\end{figure}

Example outputs of the procedure for an individual galaxy \changed{field}, NGC~2403, are shown in Figures~\ref{fig:example_spatial}--\ref{fig:example_cmd}.
NGC~2403 is a spiral in the outskirts of the M81 group, at a Galactic latitude of $b=29.2^{\circ}$ and with foreground extinction $E(B-V)=0.11$.
For comparison, \changed{I also show} the results from applying the same algorithm to a control field of the same area and at the same Galactic latitude about a degree away.
\changed{For both fields, foreground sources were identified using \autoref{eq:parallax_crit} for non-zero parallax or \autoref{eqn:pm} for significant non-zero proper motion; 
Eqns~\ref{eq:pmra} and \ref{eq:pmdec} were not used since no proper motion for NGC~2403 is available.
Results show that the proper motion criterion (Eqn.~\ref{eqn:pm}) dominates the foreground identification at all galactocentric distances: nearly all sources with significant parallaxes also have significant proper motions, but many sources have significant proper motions without significant parallax.}

\changed{The galaxy field contained 33 per cent more \gaia sources over the same area, compared to the control field, indicating that at least some \gaia sources likely belong to the galaxy.}
There are 6389 \gaia sources within a \changed{projected} radius of $a_{26}=$28\farcm2 from NGC~2403, of which \changed{4424, 4115, 3775, and 3202 are identified as foreground
at $n=2$, 3, 5, and $10\sigma$, significance, respectively. 
The control field contains 4808 sources, of which 3964, 3805, 3539, and 3019 are identified as foreground at the same significance.
To demonstrate the effect of $n$ in foreground versus background classification, the sources are separated into three groups: the most likely foreground sources (classified as foreground at $n=10$), the most likely non-foreground sources (classified as non-foreground at $n=2$), and ambiguous sources which do not fall into either of the above two groups.
The ambiguous sources make up similar fractions (19 and 20 per cent, respectively) of the total for galaxy and control field, respectively; however, the galaxy field has a greater proportion of non-foreground sources (31 per cent) than the control field (18 per cent).}

\autoref{fig:example_spatial} shows that the density of \gaia sources, and particularly the \changed{highest-confidence} non-foreground sources, is highest near the centre of the galaxy \changed{field} and declines toward its outskirts.
No similar central concentration is observed for the control field.
These distributions are again consistent with the expectation that at least some \gaia sources belong to the target galaxy.

\autoref{fig:example_radial} shows the spatial distribution of \gaia sources in NGC~2403 and the control field, marginalised to only the radial distribution.
This representation more clearly shows the decline in total source density, and the increase in the proportion of sources identified as foreground, with radius. 
\changed{
Sources categorised as ambiguous show a density profile intermediate between the foreground and non-foreground sources, indicating that this category is a mixture of the other two.
The increased density of foreground and ambiguous sources in the innermost bin is unphysical and is likely due to the limitations of the procedure and/or the \gaia data in crowded regions.
(When binned by galactocentric distance, sources with $R_{gc}<0.1 a_{26}$ have 
mean uncertainties in both proper motion and parallax measurements about 50 per cent higher than those in the galaxy outskirts.)
}
As expected, the projected density of control field sources \changed{shows no concentration} toward the centre of the field.

Finally, \autoref{fig:example_cmd} shows the distribution of the different source categories in colour-magnitude space. 
Most of the brightest sources are foreground. 
The fraction of non-foreground sources increases at fainter magnitudes and toward bluer colours, \changed{and the ambiguous sources are again located between the foreground and non-foreground sources. The foreground sources with $R_{gc}<0.1 a_{26}$  are intermediate in colour and magnitude between foreground sources beyond this projected radius and non-foreground sources.}
The control field colour-magnitude distribution is similar \changed{to the galaxy field distribution} overall; it contains fewer faint, blue sources than the galaxy field, indicating that such sources are likely associated with the galaxy.

\section{Results}

\subsection{\gaia source numbers and densities per galaxy field}

\begin{figure}
\includegraphics[width=\columnwidth]{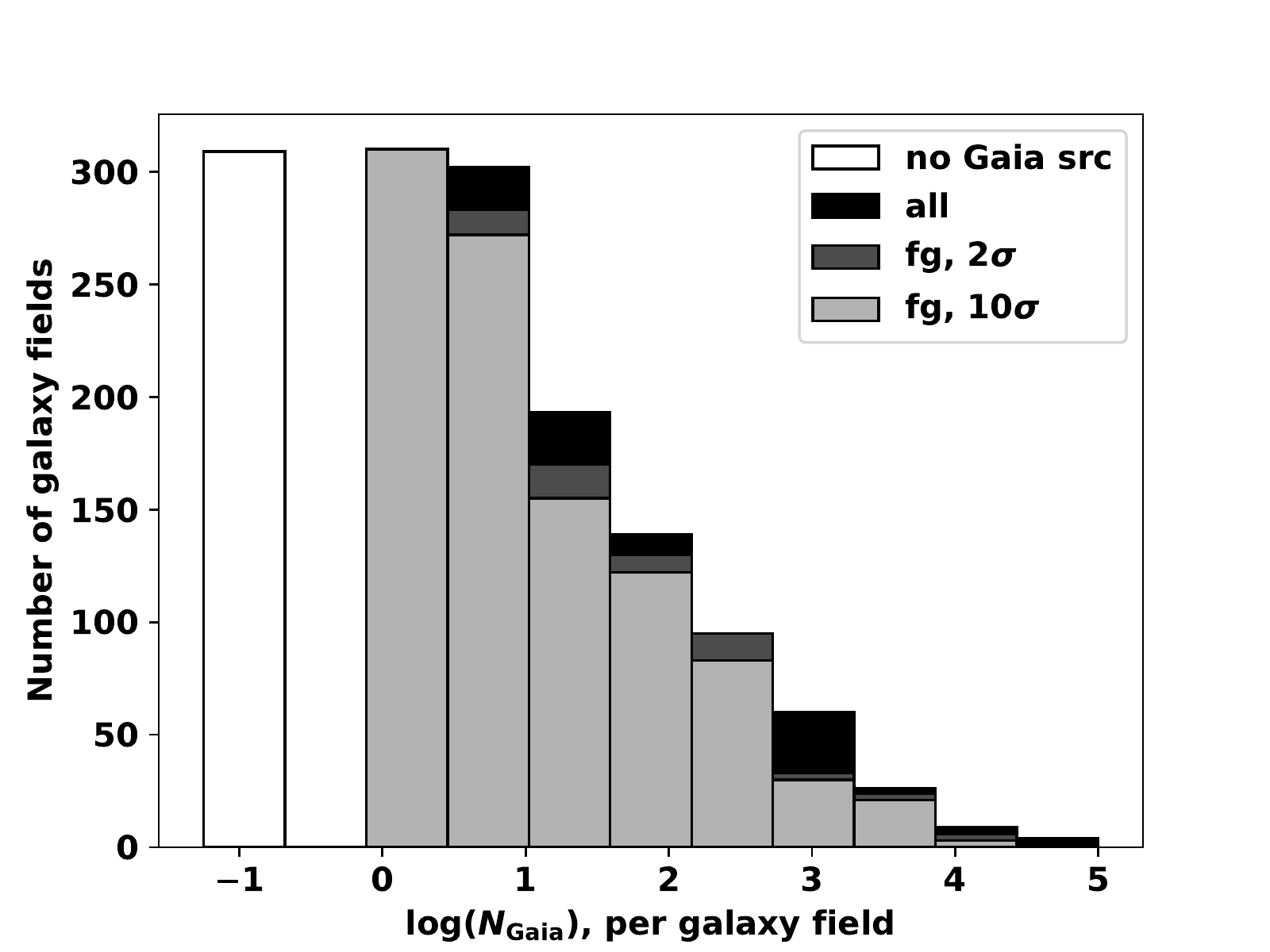}
\caption{Total number of \gaia DR3 sources (black) and foreground sources \changed{at 2 and $10\sigma$ significance per galaxy field. 
The number of galaxy fields with no \gaia sources is shown at position $\log(N_{\rm Gaia}) = -1$.}}
\label{fig:gaia_src}
\end{figure}

\changed{451422} \gaia DR3 sources are detected \changed{within a radial distance $r< a_{26}$ specified individually
for} each of the \changed{1401} galaxies in the sample.
Of these sources, a small fraction (0.6 per cent) are duplicates that lie within the search cones of multiple galaxies.
These duplicates are spread over 39 pairs of parent galaxies and as such are not expected to strongly affect the results. 
They were retained in the sample.
\autoref{fig:gaia_src} shows that the distribution of $N_{\rm Gaia}$ approximately follows a power law: most \changed{galaxy fields} have only a few \gaia DR3 sources, while a handful have thousands. 
\changed{A total of 309 galaxy field} positions have no \gaia sources at all, either foreground or non-foreground.
\changed{A similar number (298) of galaxy fields} have 50 or more \gaia sources, the minimum set for comparing the distribution of foreground and total \gaia sources with galactocentric distance or magnitude within an individual galaxy \changed{field}.

All \changed{galaxy fields} except six%
\footnote{The reader is reminded that 10 of the largest galaxies are not included here.} 
have $N_{\rm Gaia} < 12 000$, and the remainder have $N_{\rm Gaia}> 20 000$.
Listed in descending order by number of \gaia sources, these six galaxies are NGC~4945, NGC~5128, the Fornax dSph,  ESO~274-001, Circinus, and NGC~6822. 
Fornax is a very nearby ($d=140$~kpc) Milky Way satellite, NGC~6822 is a Local Group dwarf irregular  ($d=520$~kpc), and the other four objects are much more distant galaxies (3--4~Mpc) seen at low Galactic latitude. 
NGC~5128 is a giant elliptical while NGC~4945, Circinus, and ESO 274-001 are edge-on disk galaxies. 
Although five of the six (excluding Circinus) have large angular diameters ($a_{26}\gtrsim 15$~arcmin), they are by no means the only large galaxies in the LVG sample: \changed{37} galaxies fulfil this size criterion.%
\footnote{The size of M31 dwarf companion And~XIX as listed in the LVG is anomalously large at 28\farcm4. 
That this is likely one of the galaxies discussed by \citet{karachentsev2013} as a ``dwarf extremely low surface brightness [where] the diameter $a_{26}$ instead corresponds to the exponential scale $h$ of [the] brightness profile.''}

\begin{figure}
\includegraphics[width=\columnwidth]{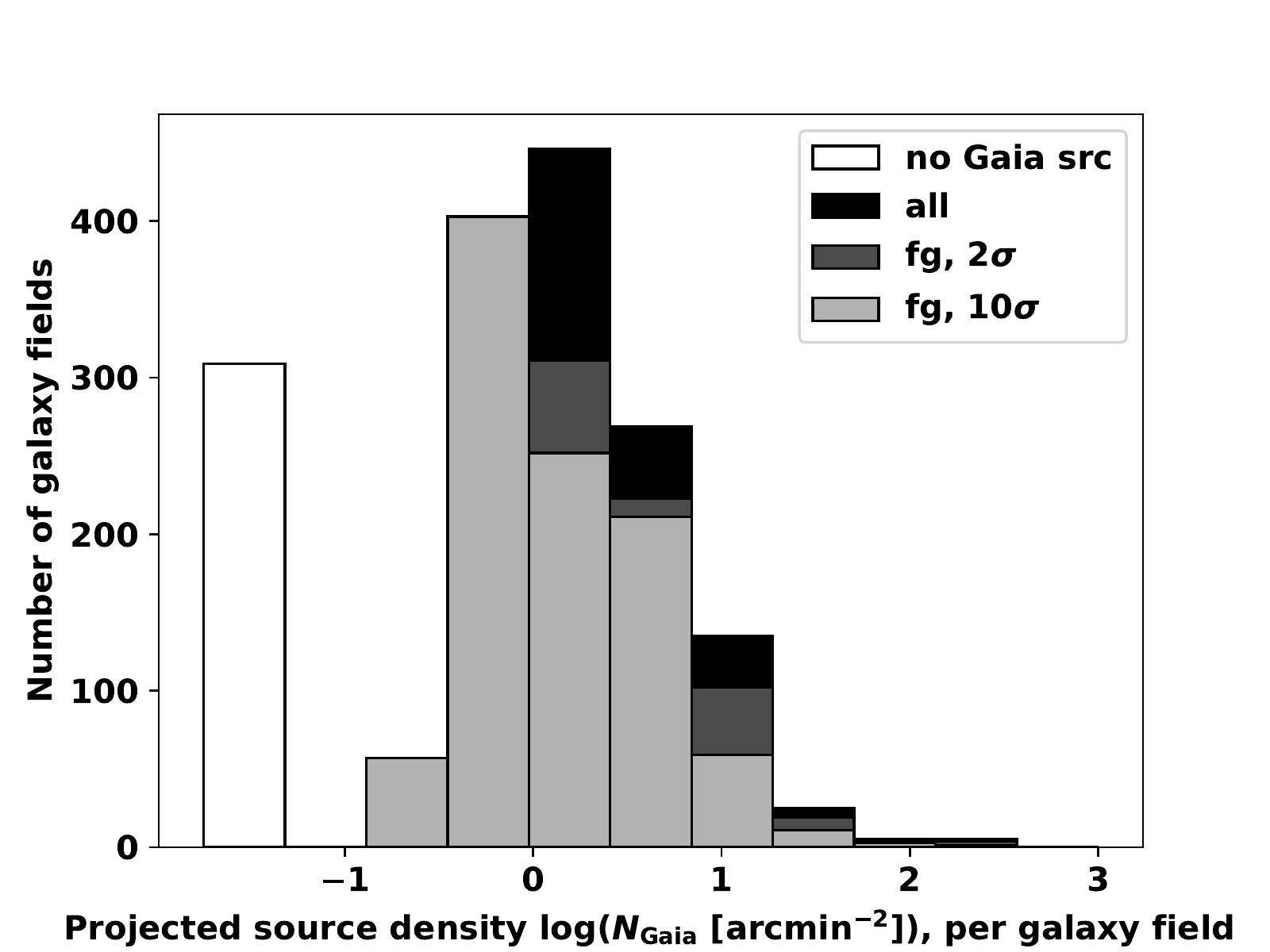}
\caption{Distribution of projected number density $\Sigma$ of \gaia sources (total number divided by area of search cone) 
\changed{at 2 and $10\sigma$ significance per galaxy field. 
The number of galaxy fields with no \gaia sources is shown at a density value of $\log(\Sigma) = -1.5$.}
}
\label{fig:number_dens1}
\end{figure}

\begin{figure}
\includegraphics[width=\columnwidth]{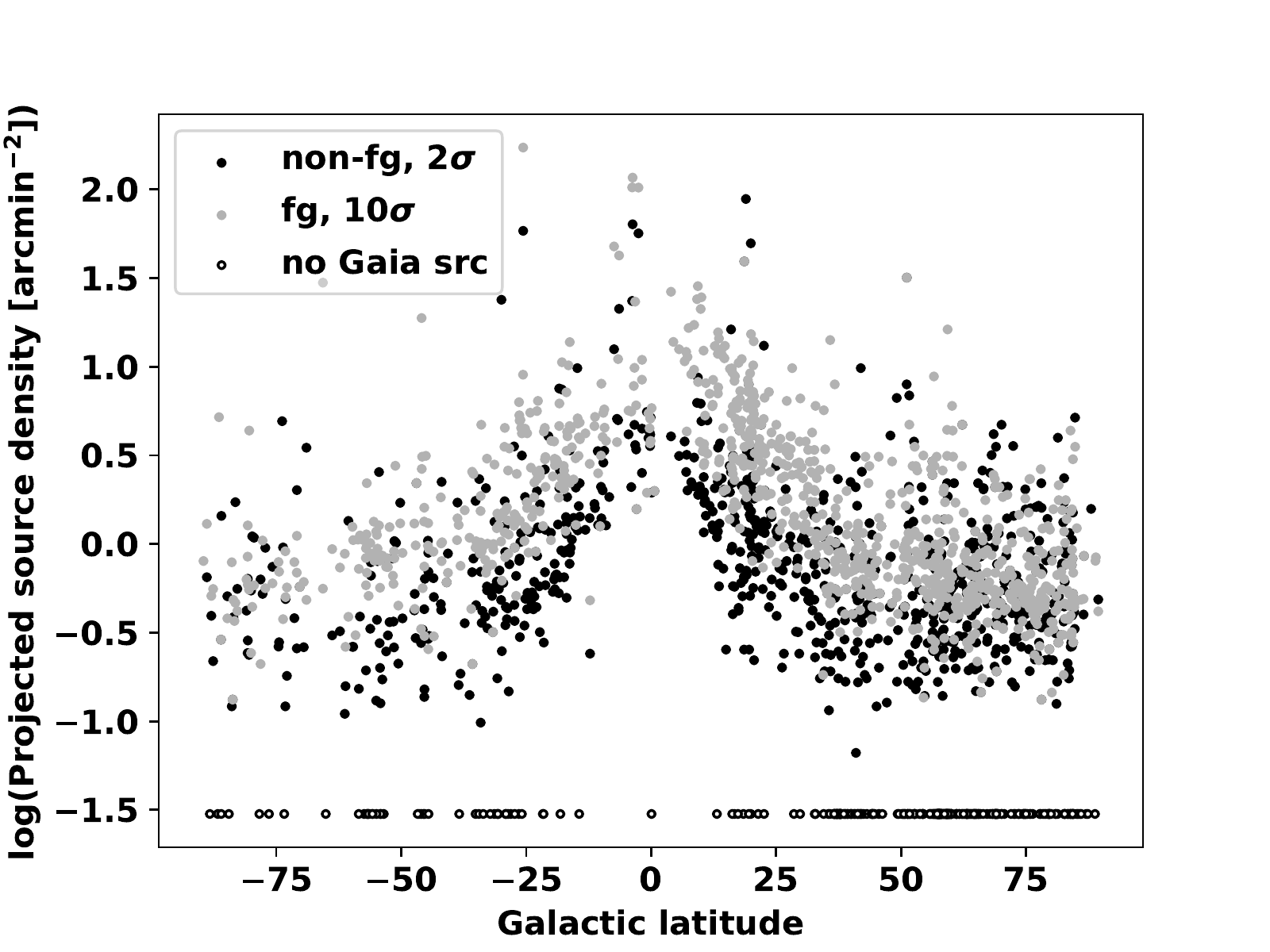}
\caption{Projected number density $\Sigma$ of \gaia sources (foreground and non-foreground) as a function of Galactic latitude.
\changed{Distribution of galaxy fields with no \gaia sources is shown at a density value of $\log(\Sigma) = -1.5$.}
}
\label{fig:number_dens2}
\end{figure}

As expected, galaxies that subtend a larger area on the sky often have more \gaia sources detected: both foreground and non-foreground sources should increase with area. 
This can be accounted for by computing the average projected source density as $\Sigma = N_{\rm Gaia}/\pi a_{26}^2$.
As \autoref{fig:number_dens1} shows, the projected source density varies by about three orders of magnitude between galaxies, so galaxy size is not the only driver of \gaia source numbers.
There is a break in the projected source density distribution at $\Sigma\sim25$~arcmin$^{-2}$: galaxies above this break with $N_{\rm Gaia}>50$ are Fornax (for total sources but not foreground sources), Circinus, and ESO~274-001, discussed above, and \changed{ten} faint irregular galaxies at $|b|<10^{\circ}$ that are foreground-dominated \changed{or that may have poor size measurements (ESO~273-014, ESO~223-009, ESO~137-018, RKK1610, EZOA J2120+45, [KK2000] 59, HIPASS J1441-62, HIZOA J1616-55, Bedin I)}.

Some but not all of the projected density variation \changed{between galaxy fields} is due to Galactic foreground.
\autoref{fig:number_dens2} shows that while there is a general trend that the minimum source density per galaxy \changed{field} increases at lower Galactic latitude, 
many low-latitude galaxies have few \gaia sources, and some high-latitude nearby galaxies have many.
At high Galactic latitudes ($|b|>25^{\circ}$) the relationship between $b$ and projected source density is weak, with a scatter of about 1~dex.
\changed{Non-foreground source density might be expected to be independent of galactic latitude or even to decrease towards low $|b|$; however, \autoref{fig:number_dens2} does not confirm these expectations. 
This implies that some sources identified here as non-foreground, even with a low threshold for significant proper motions, likely belong to the foreground. This incompleteness is quantified using spatial distributions in the following subsection.}

\changed{
The classification of \gaia sources provided by \citet{delchambre2022} provides a check on the quality of this work's classification based on astrometric criteria.
Briefly, \citet{delchambre2022} classified \gaia sources as stars, galaxies, quasars, white dwarfs or physical binary stars, using a combination of astrometric, photometric, and position information. 
Extragalactic stars or other components of nearby galaxies are not a focus of that investigation, and indeed those authors remark that sources projected near the Magellanic Clouds suffer significant misclassification.
A detailed study of these classifications is the subject of future work (Hales \& Barmby, in prep.); as a first comparison to the astrometric classification, I use the DR3 entry \textcode{classprob\_dsc\_combmod} entry and assign each source the classification with the highest combined probability above 50 per cent, and consider the foreground/non-foreground separation at $3\sigma$ significance.
Stars vastly outnumber all other categories in the \citet{delchambre2022} classification: overall, 92 per cent of \gaia DR3 sources in the vicinity of nearby galaxies (97 per cent of foreground sources and 84 per cent of non-foreground sources) are classified as stars.
As \citet{delchambre2022} point out, such a large class imbalance complicates accuracy computations, since classifying every source as a star would result in a reasonably high accuracy.
Considering the extragalactic sources in the vicinity of nearby galaxies as classified by \citet{delchambre2022}, 91 per cent of the 9760 quasars and 80 per cent of the 7028 galaxies were classified as non-foreground by the astrometric criteria in Eqns.~\ref{eq:parallax_crit}--\ref{eq:pmdec}.
This comparison shows that the astrometric criteria are producing results consistent with the more complex classification possible from using more \gaia measurements.}

\subsection{Foreground and non-foreground \gaia sources: spatial distributions}

Over the nearly half a million \gaia sources detected in the vicinity of nearby galaxies, \changed{65.5 per cent} are identified as foreground stars by applying the algorithm above to DR3 parallax or proper motion measurements.%
\footnote{In the remainder of this analysis we consider $n=3$ in \autoref{eq:parallax_crit}--\ref{eq:pmdec} as the classification criterion.}
Again this fraction varies for individual galaxies, with the average foreground fraction per galaxy \changed{field} (excluding galaxy \changed{fields} with zero \gaia sources) being \changed{69.7 per cent}.
Of the galaxies with a large number of \gaia sources discussed above, Fornax is unusual in having a very low foreground fraction ($\sim4$ per cent), while roughly half of the NGC~6822 sources are foreground. 
More than 80 per cent of the \gaia sources projected near NGC~5128, NGC~4945, Circinus, and ESO~274-001 are  foreground; these galaxies have among the highest foreground fractions of any in the LVG sample.
There are no strong trends in fraction of \gaia sources identified as foreground with either galaxy size or distance.

\begin{figure}
\includegraphics[width=\columnwidth]{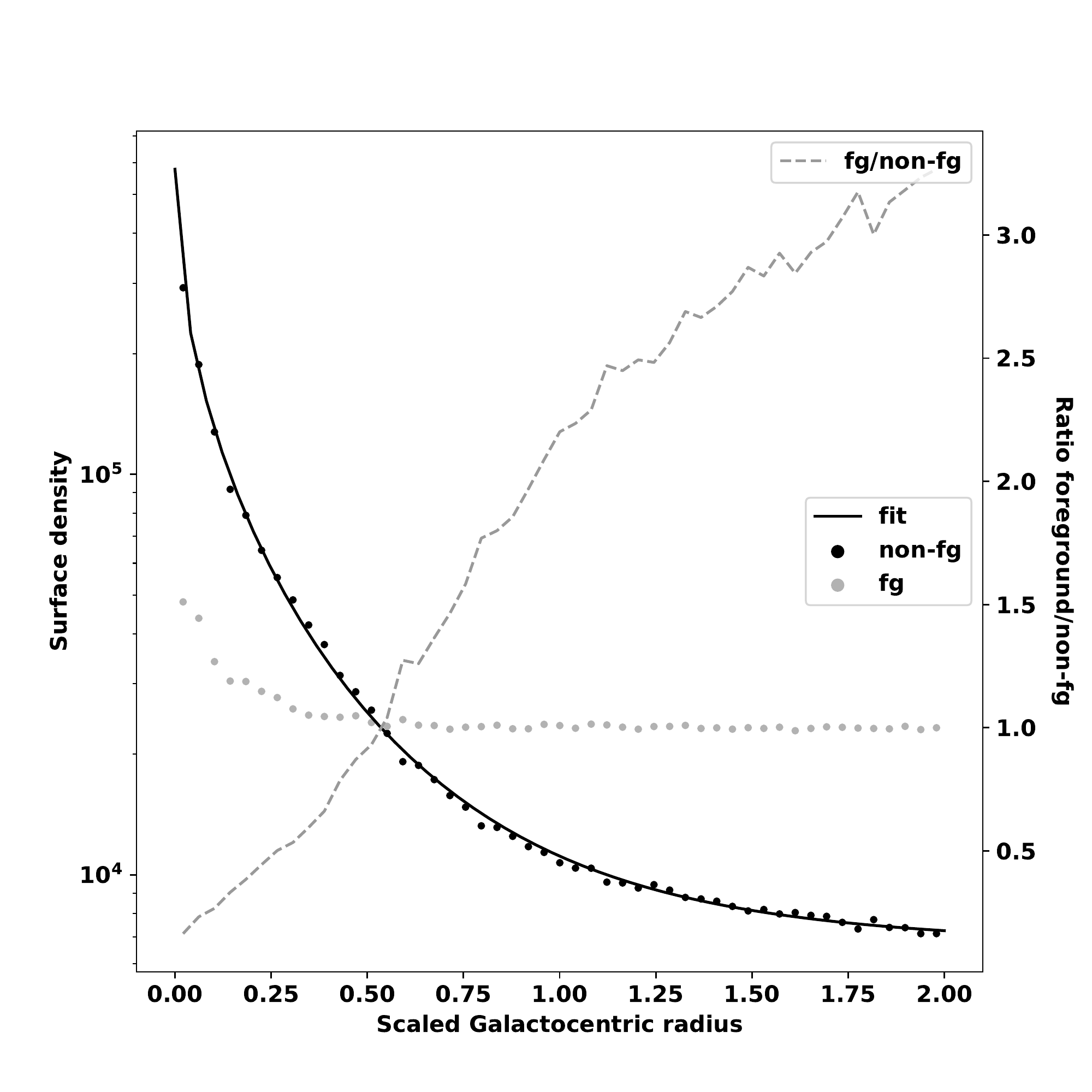}
\caption{Surface density as a function of scaled galactocentric radius ($\tilde{r} = r_{\rm GC}/(2 a_{26})$) for foreground and non-foreground \gaia sources in the vicinity of nearby galaxies.
Dashed line shows the ratio of foreground to non-foreground density.} 
\label{fig:spatial_dist}
\end{figure}

Where within a galaxy are foreground sources important? 
\autoref{fig:spatial_dist} shows the surface density of foreground- and non-foreground sources within galaxies, over the full sample. 
To facilitate comparison between galaxies of different physical and angular sizes, the projected galactocentric distance of each source has been scaled by the isophotal radius of its parent galaxy as $\tilde{r} = r_{\rm GC}/(2 a_{26})$. 
A Komolgorov-Smirnov  test finds that the two distributions are significantly different, with median values of $\tilde{r} = 1.00$ for non-foreground sources and \changed{1.40} for foreground sources. 
The ratio of foreground to non-foreground source projected densities is about \changed{0.21} closest to the galaxy \changed{field} centres and approximately \changed{3.2} at $\tilde{r} = 2$, with foreground and non-foreground sources having equal densities at ${\tilde{r}}_{50} = 0.54$.

As expected, the surface density of foreground sources is nearly constant with projected radius, giving confidence that the method is correctly identifying them.
The projected density of non-foreground sources  falls off with increasing radius and is well-fit by the combination of a S\'ersic profile $\Sigma \propto \exp (-\tilde{r}^{1/n})$ with \changed{$n=1.97$} plus a constant background density. 
Given that this density profile is the result of amalgamating data from many galaxies, the exact value of $n$ is not physically meaningful; however the fact that it lies between the $n=1$ profile of disc galaxies and the $n=4$ profile for bulge dominated systems gives confidence that these sources are indeed associated with the galaxies.
In this fit the constant density term (representing foreground and background sources not identified as such) is \changed{1.2} per cent of the central density, and \changed{26} per cent of the asymptotic projected density reached by the identified foreground sources. 
\changed{At large radii the density of sources belonging to the parent galaxy should tend to zero, so the constant density term comprises background sources and foreground sources not identified as such. }
Thus we can estimate that the \gaia sources identified as foreground represent about three quarters of the true number of {\changed contaminating sources, to the \gaia magnitude limit.}

For the \changed{298} galaxies with more than 50 \gaia sources,  ${\tilde{r}}_{50}$ was computed individually.
There is considerable variation in this quantity, with no obvious trends with galaxy distance, angular size, morphological type or luminosity.
Although ${\tilde{r}}_{50}$ has considerable scatter, it does increase with distance from the Galactic plane. 
The data are reasonably well fit by a linear trend \changed{${\tilde{r}}_{50} = 0.226+0.006|b|$}, consistent with the notion that the surface densities of foreground stars decreases as $|b|$ increases.
The median value of ${\tilde{r}}_{50}$ is \changed{0.56} with a standard deviation of \changed{0.33}, and a range from \changed{0.06 for NGC~5206 ($a_{26}$=3\farcm72, 643 foreground sources of 759 total) to 1.91 for the dwarf irregular Phoenix ($a_{26}$=5\farcm5, 69 foreground sources of 434 total)}.
Phoenix is a Milky Way satellite at high Galactic latitude, whose value of $a_{26}=$5\farcm5 as listed in the LVG may be an underestimate (e.g. the NASA Extragalactic Database gives 7\farcm7);
if its true size is 50 per cent larger, the  ${\tilde{r}}_{50}$ value would be correspondingly reduced.

\subsection{Foreground and non-foreground \gaia sources: colour and magnitude distributions}

\begin{figure}
\includegraphics[width=\columnwidth]{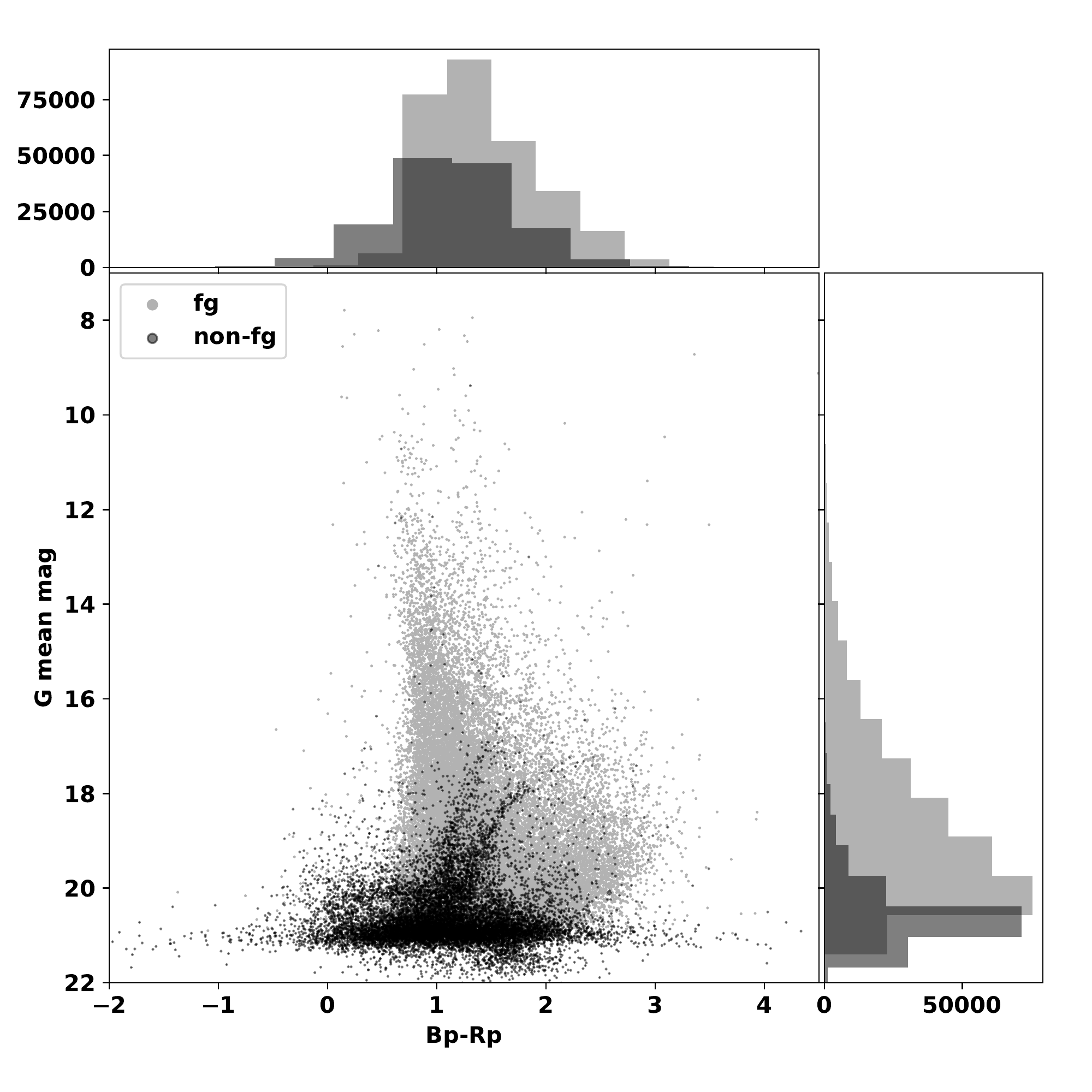}
\caption{Joint colour-apparent magnitude distribution for foreground and non-foreground \gaia sources in the vicinity of nearby galaxies. 
Histograms show all objects while the CMD plots only every tenth point.
Measurements are not corrected for extinction.} 
\label{fig:cmd_dist}
\end{figure}

\autoref{fig:cmd_dist} shows where in the colour-magnitude diagram (CMD) foreground sources are important.
Foreground sources completely dominate the brightest \gaia sources, outnumbering non-foreground sources by a factor of $\gtrsim100$ at $G<15$. 
Here the limitations of the \gaia astrometric data are apparent: there are essentially no \changed{identified} foreground sources with $G>21$.
The two populations reach equal numbers at $G\approx 20.6$, and non-foreground sources outnumber foreground sources by a factor of about 50 at the faintest magnitudes.
However, even when restricting the non-foreground source sample to the magnitude limit \changed{at which foreground identification stops} ($G=21.1$), the distributions of the two populations are still significantly different as indicated by a KS test: the foreground sources are redder and brighter than the non-foreground sources (mean $B_p-R_p=1.43$ and 1.11; mean $G=18.67$ and 20.46, respectively). 
This is compatible with the inference that the foreground sources are primarily nearby dwarfs while the non-foreground sources are more intrinsically luminous giants, supergiants, or star clusters.
\changed{The colour distributions of the two populations are much more similar than their magnitude distributions. 
There are about twice as many foreground as non-foreground sources in the central colour peak at $B_p-R_p\approx1.25$, with non-foreground sources outnumbering foreground sources only on the blue side of this peak.}

\begin{figure}
\includegraphics[width=\columnwidth]{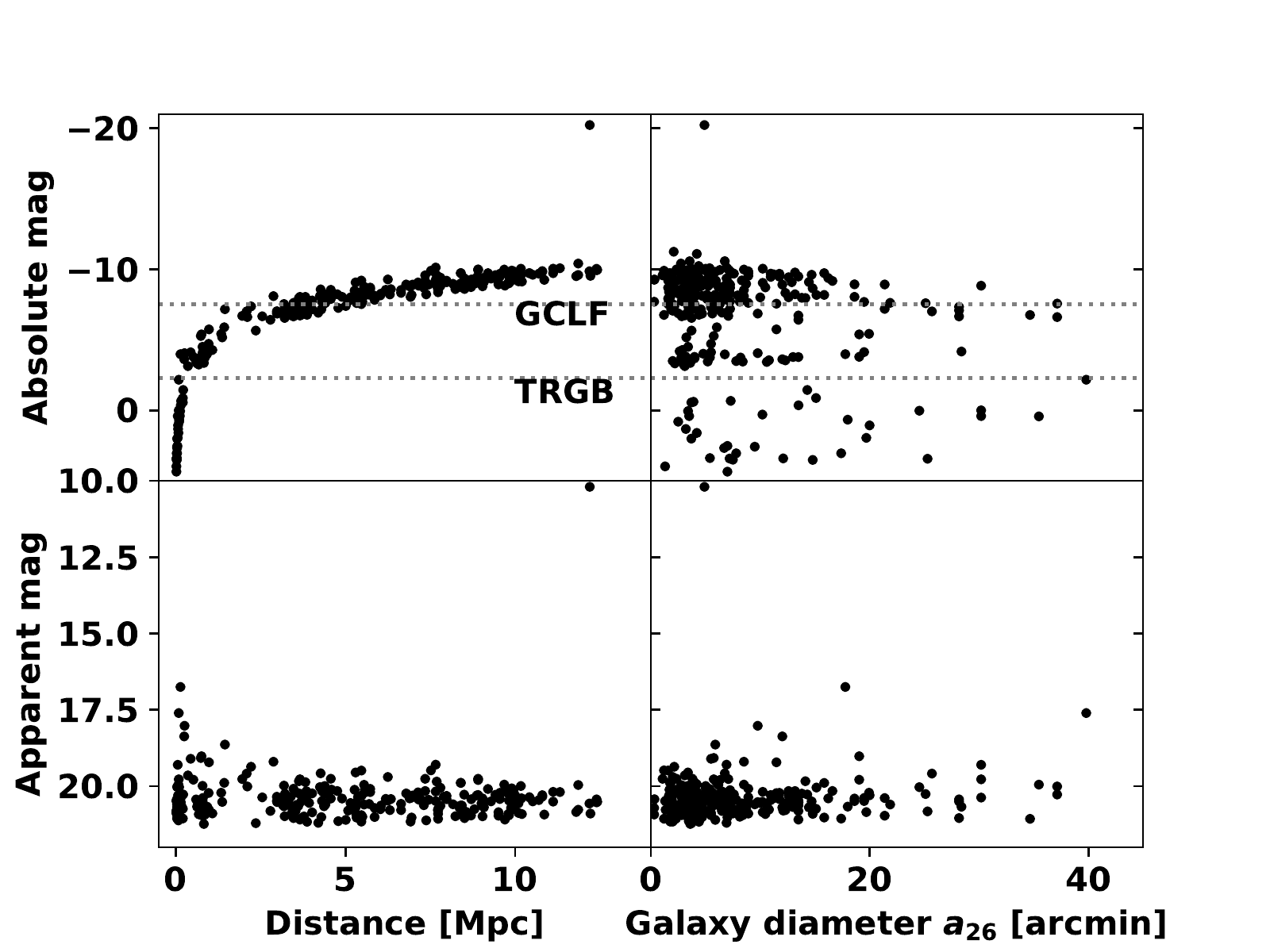}
\caption{Apparent and absolute $G$ magnitudes at which foreground sources comprise 50 per cent of the total number of \gaia sources within an individual galaxy \changed{field}, shown as a function of galaxy distance (left) and size (right). Typical values for tip of the red giant branch (TRGB) and globular cluster luminosity function peak  (GCLF) are indicated.
} 
\label{fig:absmag_dist}
\end{figure}

For the \changed{298} galaxies with more than 50 \gaia sources, the crossover magnitude, at which foreground and non-foreground source numbers are equal, was computed individually.
\autoref{fig:absmag_dist} shows these values in both apparent magnitude and absolute magnitude at an individual galaxy's distance using the distances given in the LVG. 
A few of the nearest galaxies have apparent $m_{50} <19$ but the remainder all scatter around the median value of $m_{50} = 20.50$ (standard deviation 0.82), with no obvious trends with galaxy distance, angular size, morphological type or luminosity. 
There are also no clear trends with Galactic latitude or foreground extinction.
Because of the steep increase in the faint end of the magnitude distribution (see \autoref{fig:cmd_dist}), the kernel density estimates of $m_{50}$ can be strongly driven by a few outliers for galaxies with a small number of sources.
For example, IC~2233 and Segue~1 each have $<200$ sources, with $\sim80$ per cent being foreground and only a few non-foreground sources brighter than $G=20$, yet IC~2233 had the brightest apparent $m_{50}$ ($10.20$) and Segue~I the faintest \changed{(20.86)}.
With a much larger number of sources (\changed{1434} foreground sources of 33433 total), Fornax has the second-brightest $m_{50}=16.75$.
The absolute magnitude $M_{50}$ values are strongly dependent on distance, indicating that the foreground-to-non-foreground magnitude crossover points are primarily driven by the limits of the \gaia data.

For an individual galaxy \changed{field}, it is also important to compare the apparent magnitudes of foreground sources to those of relevant features in the galaxy's stellar population such as the tip of the red giant branch \citep[TRGB:  $M_G=-2.3$,][]{soltis2021} or the peak of the globular cluster luminosity function \citep[GCLF: $M_G=-7.5$,][]{rejkuba2012}.
\autoref{fig:absmag_dist} shows that the current \gaia limit for identifying foreground sources ($G\approx 21$) corresponds to the absolute magnitudes of the TRGB ($G=-2.3$) at a distance of 450~kpc, the faint end of the GCLF (2~mag fainter than the peak) at $D=2$~Mpc, and the GCLF peak ($G=-7.5$) at 5~Mpc.
Detecting foreground contamination of these features will be incomplete for galaxies beyond these distances.

\section{Discussion}

\changed{
\autoref{fig:example_cmd} and \autoref{fig:cmd_dist} both show sharp cutoffs in the colour of foreground stars at $B_p-R_p\approx0.75$. Although this feature is visible in other similar figures made with Gaia photometry \citep[e.g.][]{fouesneau2022, van_der_marel2019}, I was unable to find it discussed in the literature. 
By comparison with fig.~1 of \citet{daltio2021} and fig.~1 of \citet{bailerjones2022}, it can be inferred that this colour corresponds to both the upper main sequence turn off in the solar neighbourhood and the approximate blue limit of galaxy colours.
Varying distances and luminosities cause objects of this colour to spread in apparent magnitude.
}

Limitations of this work include the fact that
many \gaia sources are too faint for reliable proper motion and/or parallax determinations in the current data release; future data releases can be expected to include more likely foreground sources. 
Improvements in \gaia data processing may lead to improved measurements in crowded regions such as near the centres of galaxies, although fundamental limits of spatial resolution will remain. 
\changed{Future development of a probabilistic classification method that makes use of uncertainties in the astrometric quantities is one way to better quantify these limits.}
Foreground sources whose space motion is primarily along, rather than tangential to, the line of sight will not be easily distinguishable via proper motions. 
Radial velocities and other \gaia measurements beyond parallax and proper motion may potentially be useful in future work.
In this work no attempt was made to distinguish between \gaia sources within a target nearby galaxy and true background sources.
For target galaxies at greater distances, the increase in number counts of background galaxies with magnitude means that contamination from background sources is expected to be more severe.
The distances to the target galaxies are not always well-determined; however galaxy distances only come into play when considering colour-absolute magnitude diagrams.

\section{Conclusions}

Studies of resolved stellar populations in nearby galaxies are contaminated by both foreground sources (Milky Way stars) and background sources (galaxies and active galactic nuclei). 
Averaged over a sample of \changed{1401} nearby galaxies, about two thirds of \gaia DR3 sources near these galaxies can be identified as foreground from their proper motion and/or parallax measurements.
The fraction of foreground sources is anticipated to increase in future \gaia data releases, as astrometric solutions are measured for fainter sources.
The foreground sources are redder and brighter than non-foreground sources. 

As might be expected, foreground sources dominate near the outskirts of galaxies and at brighter magnitudes.
On average, foreground sources outnumber non-foreground sources at galactocentric radii \changed{$r>1.06 a_{26}$} and at apparent magnitudes \changed{$m_{G}<20.50$}. 
Considered on an individual galaxy \changed{field} basis, the radius where foreground sources outnumber non-foreground sources  shows substantial variation, with no clear dependence on galaxy properties.
The apparent magnitude at which foreground sources outnumber non-foreground sources is much less variable across \changed{galaxy fields}, and is primarily due to the \gaia signal-to-noise limits. 
This means that the depth reached by \gaia studies of stellar populations in galaxies will be limited by galaxy distances: 
galaxies within 450~kpc have secure foreground identification at the absolute magnitude of the  tip of the red giant branch, while galaxies within 2~Mpc
have foreground identification two magnitudes beyond the globular cluster luminosity function peak.
For galaxies at distances beyond 5~Mpc, \gaia\ foreground removal for these features will be incomplete.

\gaia DR3 data provide a useful avenue for removing foreground contamination in photometric studies of stellar populations of nearby galaxies.
The depth and uniformity of \gaia data will no doubt facilitate many investigations of stars and star clusters in the nearby Universe.

\section*{Acknowledgements}
I thank the referee for helpful comments that improved the paper and A. Hughes, M. Gorski and J. Hales for helpful discussions. 

This work has made use of data from the European Space Agency (ESA) mission
{\it Gaia} (\url{https://www.cosmos.esa.int/gaia}), processed by the {\it Gaia}
Data Processing and Analysis Consortium (DPAC,
\url{https://www.cosmos.esa.int/web/gaia/dpac/consortium}). Funding for the DPAC 
has been provided by national institutions, in particular the institutions
participating in the {\it Gaia} Multilateral Agreement.

\section*{Data Availability}

The data underlying this article are available in Zenodo, at {\url{https://doi.org/10.5281/zenodo.7244987}}.
The datasets were derived from sources in the public domain: 
\gaia  Data Release 3
{\url{https://www.cosmos.esa.int/web/gaia/data-release-3}}, 
and the Local Volume Galaxy catalogue {\url{http://www.sao.ru/lv/lvgdb/}}.


\bibliographystyle{mnras}
\bibliography{Gaia_fg}






\bsp	
\label{lastpage}
\end{document}